%% file: JAJ.tex
\documentclass[%
 reprint,
 superscriptaddress,preprintnumbers,
 amsmath,amssymb,
 aps,
]{revtex4-1}

\usepackage{graphicx}
 \usepackage{dcolumn}
\usepackage{bm}
\usepackage{color}
\usepackage[normalem]{ulem}

\usepackage{hyperref}

\begin{document}

\title{The Mono-Tau Menace: From $B$ Decays to High-$p_T$ Tails}

\author{Admir Greljo}
\affiliation{Theoretical Physics Department, CERN,
1 Esplanade des Particules, 1211 Geneva 23, Switzerland}
\author{Jorge Martin Camalich}
\affiliation{Theoretical Physics Department, CERN,
1 Esplanade des Particules, 1211 Geneva 23, Switzerland}
\affiliation{%
Instituto de Astrof\'isica de Canarias, C/ V\'ia L\'actea, s/n
E38205 - La Laguna, Tenerife, Spain
}%
\affiliation{%
Universidad de La Laguna, Departamento de Astrof\'isica, La Laguna, Tenerife, Spain
}%
\author{Jos\'e David Ruiz-\'Alvarez}
\affiliation{Instituto de F\'isica, Universidad de Antioquia, A.A. 1226, Medell\'in, Colombia}

\preprint{CERN-TH-2018-243}
\begin{abstract}

We investigate the crossing-symmetry relation between $b\to c\tau^-\bar\nu$ decay and $b\bar c\to \tau^-\bar\nu$ scattering to derive direct correlations of New Physics in semi-tauonic $B$-meson decays and the mono-tau signature at the LHC ($pp\to\tau_h X$ + MET). Using an exhaustive set of effective operators and heavy mediators we find that the current ATLAS and CMS data constrain scenarios addressing anomalies in $B$-decays. Pure tensor solutions, completed by leptoquark, and right-handed solutions, completed by $W^\prime_R$ or leptoquark, are challenged by our analysis. Furthermore, the sensitivity that will be achieved in the high-luminosity phase of the LHC will probe \textit{all} the possible scenarios that explain the anomalies. Finally, we note that the LHC is also competitive in the $b\to u$ transitions and bounds in some cases are currently better than those from $B$ decays.

\end{abstract}

\pacs{Valid PACS appear here}

\maketitle

\input{definitions_notes.tex}

\textit{\textbf{Introduction:}} Branching fractions of semi-tauonic $B$-meson decays, measured through the ratios $R_{D^{(*)}}=\Gamma(B\to D^{(*)}\tau\nu)/\Gamma(B\to D^{(*)}\ell\nu)$ (with $\ell=e$ or $\mu$), appear to be enhanced with respect to the Standard Model (SM) by roughly thirty percent, with a global significance of $\sim 4\sigma$~\cite{Lees:2012xj,Lees:2013uzd,Huschle:2015rga,Sato:2016svk,Aaij:2015yra,Hirose:2016wfn,Hirose:2017dxl,Aaij:2017uff,Aaij:2017deq,Aaij:2017tyk,Aoki:2016frl}. If this is due to new physics (NP), its mass scale is expected to be not far above the TeV scale (see e.g.~\cite{DiLuzio:2017chi}). The most immediate question is whether such NP is already ruled out by the existing high-$p_T$ searches and, if not, what is the roadmap for its direct discovery.

From a bottom-up perspective the NP interpretation of the $R_{D^{(*)}}$ anomalies involves two different aspects, \textit{(i)} new dynamics (i.e. degrees of freedom), and \textit{(ii)} the flavour structure. Both aspects are relevant when it comes to identifying correlated effects in other observables such as weak hadron or $\tau$ decays, electroweak precision observables and high-$p_T$ LHC signatures (see e.g.~\cite{Buttazzo:2017ixm}). 

The Lorentz structure of the effective operators that describe the effects of the hypothesized heavy mediators at low energies can be discriminated by using $b\to c \tau\nu$ decay data alone~\cite{Alonso:2016oyd,Datta:2017aue,Akeroyd:2017mhr,Tran:2018kuv,Nierste:2008qe,Fajfer:2012vx,Sakaki:2014sea,Ligeti:2016npd,Alonso:2016gym,Alonso:2017ktd,Asadi:2018sym}. On the other hand, most of flavor data is consistent with the SM, which suggests that such NP must couple mainly to the third generation of quarks and leptons~\cite{Bhattacharya:2014wla,Glashow:2014iga,Alonso:2015sja,Greljo:2015mma,Feruglio:2016gvd,Feruglio:2017rjo,Buttazzo:2017ixm,Kumar:2018kmr,Feruglio:2018fxo}. However, in general, and without the guidance of a theory of flavor, models addressing the anomalies have some freedom in the way they implement couplings in flavor space. All this complicates defining conclusive tests in other weak hadron decays or clear direct-search strategies at the LHC.

\begin{figure}[t]
\begin{tabular}{cc}
  \includegraphics[width=55mm]{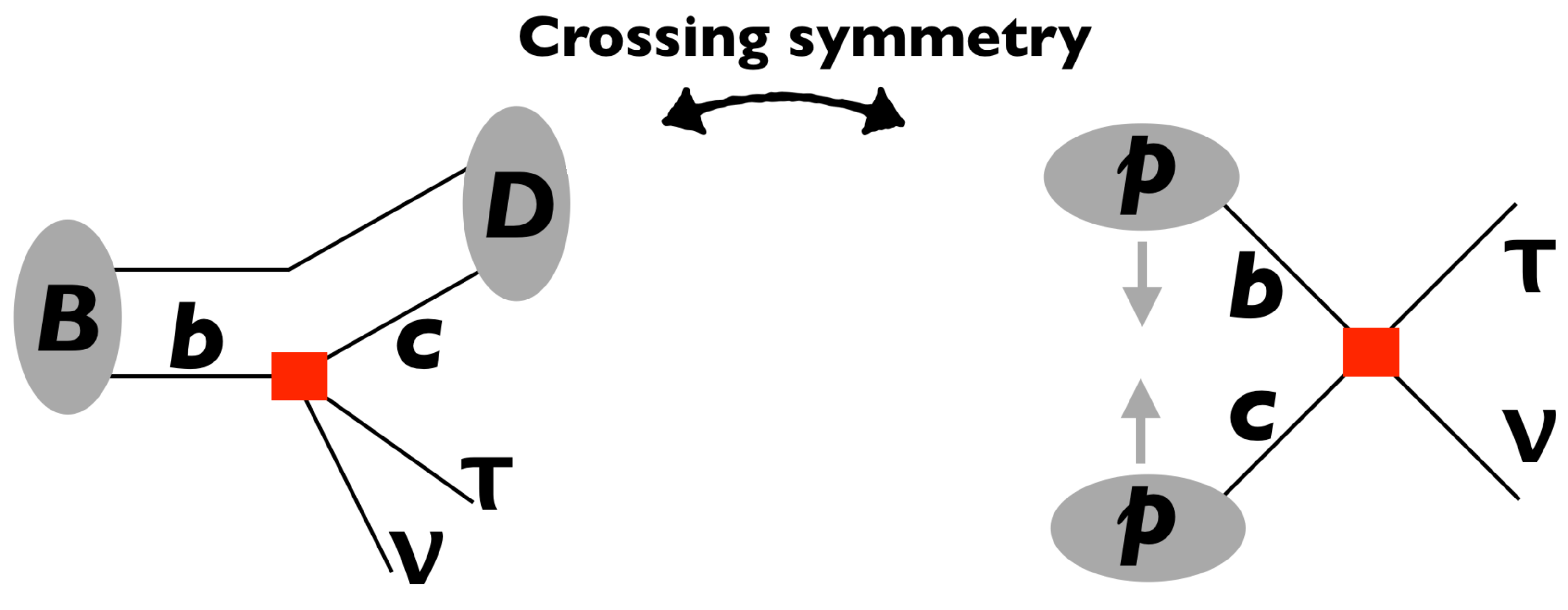} 
  \end{tabular}
\caption{Illustration of the complementarity in $b \to c \tau \nu$ transitions as measured in $B$ meson decays and inclusive production of $\tau$+MET of high-$p_T$ LHC.
\label{fig:intro}}
\end{figure}

The aim of this letter is to discuss and explore in detail the phenomenology of a collider signature that should be produced at the LHC by \textit{any} model addressing the $R_{D^{(*)}}$ anomalies with new heavy mediators. The main idea, illustrated in Fig.~\ref{fig:intro}, is that regardless of the Lorentz and flavor structure of the NP, crossing symmetry \textit{univocally} connects the $b\to c\tau^-\bar\nu$ decay and the $b \bar c \to \tau^-\bar\nu$ scattering processes~\cite{Alonso:2016oyd,Bhattacharya:2011qm,Gonzalez-Alonso:2016etj,Altmannshofer:2017poe,Abdullah:2018ets}. As we demonstrate below, the analysis of $p p \to \tau\nu X$ at the LHC already excludes broad classes of models addressing the anomalies and provides a \textit{``no-lose theorem''} for the direct discovery of NP at the LHC, in case the $R_{D^{(*)}}$ anomalies were confirmed in the future. Furthermore, these searches simultaneously constrain operators involving semi-tauonic $b\to u$ transitions with bounds that are currently competitive, or even better, than those obtained in $B$ decays. 

\textit{\textbf{Effective-field theory:}} We start with a low-energy effective field theory (EFT) of NP in semi-tauonic $b\to u_i$ transitions (with $u_i$ up- or charm-quarks)~\cite{Goldberger:1999yh,Cirigliano:2009wk},
\begin{align}
\label{eq:leff1} 
{\cal L}&_{\rm eff} \supset
- \frac{2V_{ib}}{v^2} \,
\Bigg[
\Big(1 + \epsilon_L^{ib} \Big) \bar{\tau}  \gamma_\mu P_L   \nu_{\tau} \cdot \bar u_i   \gamma^\mu P_L b\nonumber\\
&+\eR^{ib}\bar{\tau}  \gamma_\mu P_L   \nu_{\tau} \cdot \bar u_i   \gamma^\mu P_R b
+ \epsilon_T^{ib}   
\,   \bar{\tau}   \sigma_{\mu \nu}P_L \nu_{\tau}    \cdot  \bar u_i   \sigma^{\mu \nu}P_L b\nonumber\\
&+ \epsilon_{S_L}^{ib}\bar{\tau}  P_L \nu_{\tau}\cdot \bar u_iP_L b 
+ \epsilon_{S_R}^{ib}\bar{\tau}  P_L \nu_{\tau}\cdot \bar u_iP_R b
\Bigg]+{\rm h.c.}
\end{align}
where subindices label quark flavor in the mass basis, $V_{ij}$ are the Cabibbo-Kobayashi-Maskawa (CKM) matrix elements, $P_{L,R}$ are the chiral projectors, $\sigma^{\mu\nu}=i/2[\gamma^\mu,\gamma^\nu]$ and we have used 
 $v\approx246$ GeV the electroweak symmetry breaking (EWSB) scale. With this normalization, the Wilson coefficients (WCs) scale as $\epsilon_\Gamma\sim v^2/\Lambda^2$, where $\Lambda$ is the characteristic scale of  NP. Light right-handed neutrinos can be added to Eq.~(\ref{eq:leff1}) with the replacements $P_L\to P_R$ in the leptonic currents and $\epsilon_\Gamma\to\tilde\epsilon_\Gamma$ in labeling the WCs. None of these operators interfere with the SM for vanishing neutrino masses. 

In order to connect this EFT to NP with a typical scale $\Lambda\gg v$, one needs to switch first to another EFT which is invariant under $SU(2)_L\times U(1)_Y$ and is built using the full field content of the SM~\cite{Buchmuller:1985jz,Grzadkowski:2010es}.
Without specifying the flavour structure, we focus on the collider signature that stems exclusively from four-fermion operators giving $\bar c b~\bar\tau\nu$ in the fermion mass basis, which are the ones directly linked to  $R_{D^{(*)}}$. Finally, when connecting the values of the WCs at $\mu=m_b$ to those at $\mu=\Lambda$, one needs to account for the rescaling and mixing effects induced by the renormalization group evolution produced by SM radiative corrections~\cite{Jenkins:2013zja,Jenkins:2013wua,Alonso:2013hga,Aebischer:2017gaw,Gonzalez-Alonso:2017iyc,Jenkins:2017dyc}.

\begin{table}[h]
\caption{Values of the WCs at $\mu=m_b$ of the EFT Lagrangian of eq.~(\ref{eq:leff1}) for semi-tauonic $b\to c$ transitions fitted to the current values of $R_{D^{(*)}}$. For the theoretical analysis we follow ref.~\cite{Alonso:2016gym}.  \label{tab:RDsFit}}
\begin{tabular}{ccccc}
\hline\hline
Left-handed& Tensor &\multicolumn{2}{c}{Scalar-Tensor}&Right-handed\\

$\epsilon_L^{cb}$& $\epsilon_T^{cb}$&  $\epsilon_{S_L}^{cb}$&$\epsilon_T^{cb}$&$\tilde\epsilon_R^{cb}$\\
\hline
0.11(2)&0.37(1)&0.18(7)&$-0.042(10)$&0.48(6)\\
\hline\hline
\end{tabular}
\end{table}

At low energies, these operators induce semi-tauonic $B$ decays, as shown in Fig.~\ref{fig:intro} left. The characteristic $(V-A)$ structure remaining in the $\Lambda\to\infty$ limit incarnates the SM contribution, whereas different combinations of these operators have been found to accommodate the $R_{D^{(*)}}$ anomalies~\cite{Freytsis:2015qca,Bhattacharya:2018kig,Robinson:2018gza}. A sample of the preferred NP solutions is shown in Tab.~\ref{tab:RDsFit}.    

At high energies, these operators contribute to $p p \to \tau \nu X$ at the LHC, as shown in Fig.~\ref{fig:intro} right. Schematically, the ratio of NP and SM cross-sections for this process, at energies $\sqrt{s}\gg M_W$ and  leading order in QCD, reads
\begin{align}
\frac{\sigma_{\rm{NP}}}{\sigma_{\rm{SM}}} &\sim \frac{\sum_i\mathcal L_{ib}\otimes|V_{ib}|^2\frac{s}{v^4}\, \left( \alpha_\Gamma|\epsilon_\Gamma^{i b}|^2 \right)}{\mathcal L_{u d}\otimes|V_{u d}|^2\frac{s}{v^4}\, \left(\frac{M_W^2}{s}\right)^2}~,\label{eq:xsec_scheme}
\end{align}
where the sum over flavors refers to the $up$ and $charm$ quark in Eq.~\ref{eq:leff1} and are convoluted by the luminosity functions $\mathcal L_{ij}$ containing the corresponding parton distribution functions (PDF). The SM cross-section is given by the $W^\pm$ exchange while in the NP one $\alpha_\Gamma$ is an operator-dependent factor (e.g. $\alpha_L = 1$). The sensitivity to NP in $b \to u_i$ comes from the quadratic dependence on the WCs, while the interference with the SM is relevant only when involving both $up$ and $down$ flavors~\cite{Bhattacharya:2011qm,Cirigliano:2018dyk}.

At first glance, one might conclude that effects in $b \to u_i$ are negligible when compared with the dominant SM production from $u \bar d, d \bar u$ fusion, which is PDF and CKM favoured. However, in the high-$p_T$ tails above the EWSB scale, the SM amplitude unitarises while the EFT one keeps growing. Interestingly, the energy enhancement in the tails is large enough to compensate for the aforementioned suppressions leading to bounds competitive to $B$-decays. Finally, the absence of interference effects implies that the collider signature is sensitive only to the Lorentz structure (``vector'', ``scalar'' or ``tensor'') and not to the chirality of the partonic currents.

\begin{figure}[t]
\begin{tabular}{cc}
  \includegraphics[width=90mm]{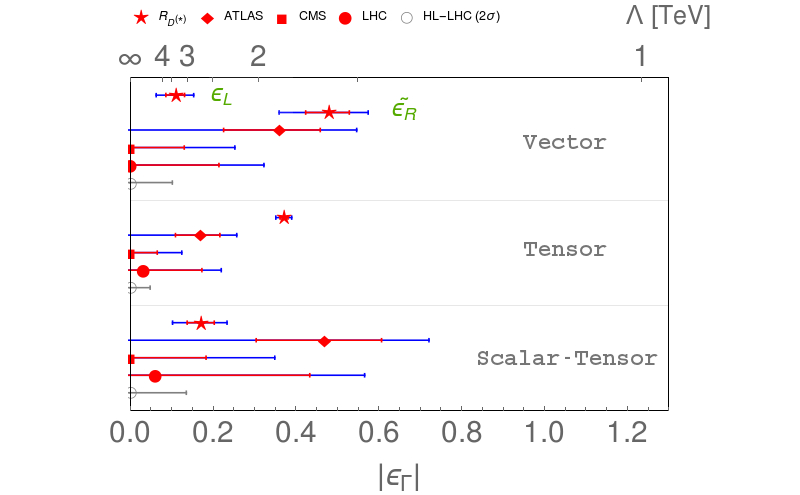}
  \end{tabular}
\caption{1$\sigma$ (red) and $2\sigma$ (blue) ranges on the absolute value of the WCs of semi-tauonic $cb$ transitions at $\mu=m_b$.~\label{fig:EFTcb}}
\end{figure}

To perform our numerical collider studies we use {\tt MadGraph5\_AMC@NLO {v2.6.1}}~\cite{Alwall:2011uj, Alwall:2014hca} with the NNPDF 3.0 PDF set (and using {\tt FeynRules 2.0}~\cite{Alloul:2013bka}) to generate samples of the inclusive process $p p \to \tau_h X+{\rm MET}$. We work at leading order in QCD but we add up to two jets at the partonic level, introducing $(\alpha_s/\pi)$-suppressed NP contributions through e.g. $g\bar c\to \bar b \tau^-\bar\nu$~\cite{Altmannshofer:2017poe} or $g g \to c\bar b \tau^-\bar \nu$ on top of the numerically more significant $b \bar c \to \tau^-\bar\nu$. The output is matched to {\tt Pythia 8 {v8.230}}~\cite{Sjostrand:2014zea} for modeling the parton showers and hadronization and, finally, to {\tt Delphes {v3.4.1}}~\cite{deFavereau:2013fsa} for proper detector simulation {with default ATLAS and CMS detectors configurations}. We compare our simulations to $W^\prime$ searches in this channel performed by ATLAS with $36.1$ fb$^{-1}$~\cite{Aaboud:2018vgh} and CMS with $35.9$ fb$^{-1}$~\cite{Sirunyan:2018lbg}. Simulations are ran independently for each experiment and we apply the same kinematic cuts described in their papers. 

A good agreement, within a $\sim 20\%$, is obtained between our simulated transverse mass distributions of the $\tau_h$ ($m_T$) in $W^-\to\tau^-\bar\nu$ and those reported by the experimental collaborations. The total signal is compared to the $m_T$ distributions measured by ATLAS and CMS assuming Poissonian probabilities for the events in each bin~\cite{Cowan:2010js}. In our analysis of NP, we systematically take into account the renormalization-group evolution of the WCs by assigning $\mu$ equal to the average $m_T$ in each bin. Systematic uncertainties of the SM backgrounds reported by the experiments are incorporated in the analysis by means of nuisance parameters that we assume to be normally distributed and uncorrelated. The results of the statistical analyses presented in this work stem from the profile likelihoods depending exclusively on the WCs. In addition to the analysis of the current data, we perform a sensitivity study for the LHC after run 2 (150 fb$^{-1}$) and after the HL-LHC phase (3 ab$^{-1}$), assuming that the systematic uncertainties of the SM background scale with luminosity as $\delta/N\sim 1/{\sqrt{N}}$~\cite{CMS:2017cwx}.

\begin{table}[h]
\caption{$2\sigma$ upper bounds for the absolute value of the WCs of semi-tauonic $cb$ transitions at $\mu=m_b$.~\label{tab:EFTcb}}
\vspace{0.2cm}
\begin{tabular}{cccc}
\hline\hline
Data set& Vector& Scalar & Tensor\\
\hline
ATLAS (36.1 fb$^{-1}$)& $0.55$&$0.93$&$0.26$\\
CMS (35.9 fb$^{-1}$) &$0.25$&$0.45$&$0.12$\\
LHC combined          & $0.32$ & $0.57$ & $0.16$\\
\hline
LHC (150 fb$^{-1}$)       &$0.21$ & $0.37$ & $0.10$\\
HL-LHC  & $0.10$ & $0.17$ & $0.05$\\
\hline\hline
\end{tabular}
\end{table}

\begin{figure*}[t]
\begin{tabular}{cc}
  \includegraphics[width=75mm]{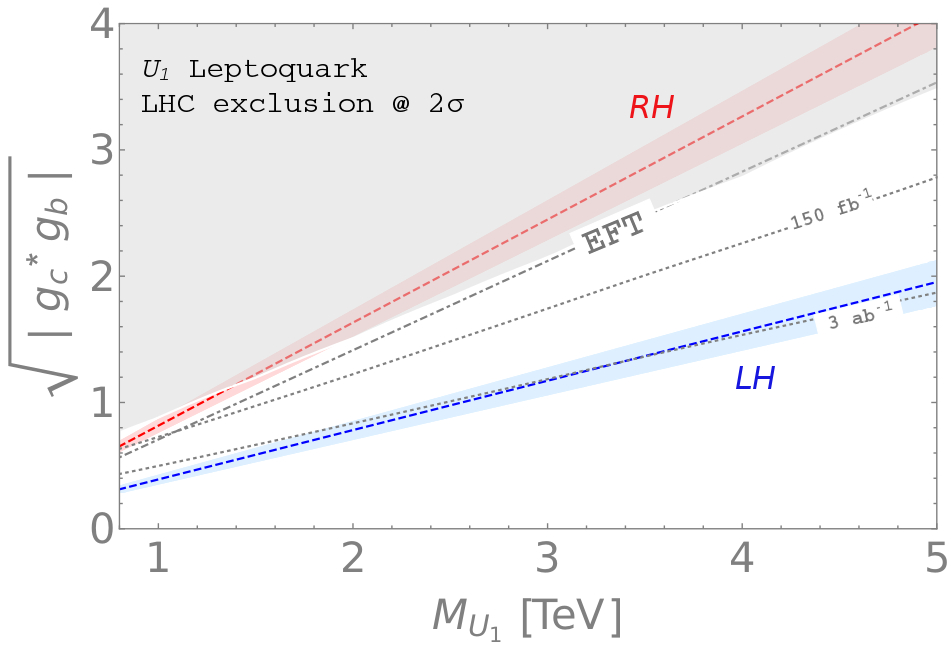}\hspace{0.5cm} & \hspace{0.5cm} \includegraphics[width=77mm]{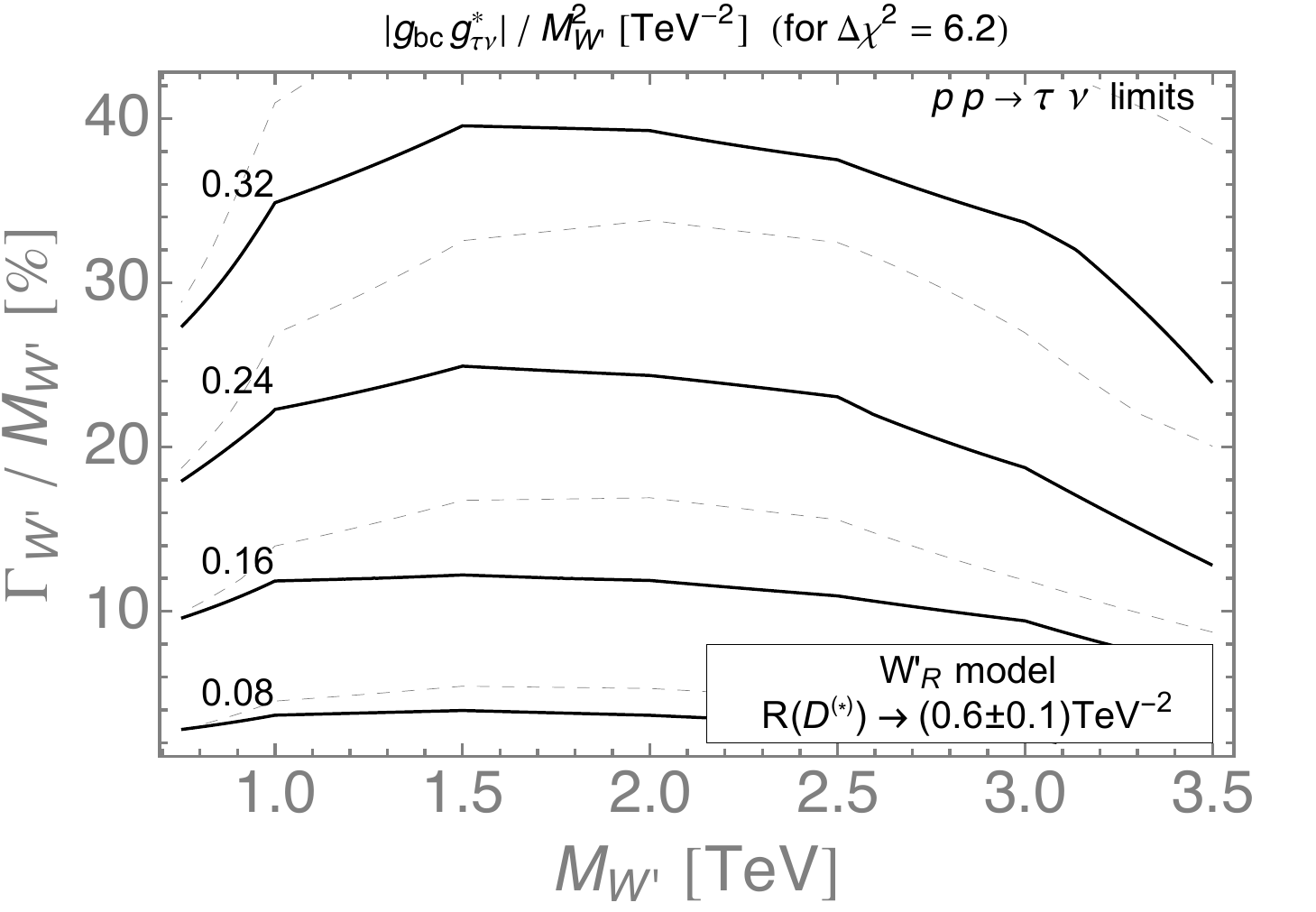}
  \end{tabular}
\caption{Bounds on representative explicit models that address the $R_{D^{(*)}}$ anomalies. \textit{Left:} The $U_1$ vector leptoquark. \textit{Right:} A potentially broad $W'$ gauge boson. See main text for details. 
\label{fig:simplified_models}}
\end{figure*}

In Tab.~\ref{tab:EFTcb} we show the results of our NP collider analysis in terms of the $cb$ four-fermion operators. The fits to the two collaborations differ mainly because ATLAS has a slight excess of events in the $m_T$ distribution whereas the one of CMS is systematically consistent with the SM. The most remarkable result shown in this table is that, combining the analysis of the two sets of data, we arrive at a sensitivity to NP which is, indeed, competitive to the one achieved in $B$ decays. In fact, the collider data poses already a challenge to some of the possible explanations to the $R_{D^{(*)}}$ anomaly. To make this discussion clearer, we compare in Fig.~\ref{fig:EFTcb} the results from the fits to $R_{D^{(*)}}$ shown in Tab.~\ref{tab:RDsFit} with the ones obtained from the collider analysis. The tensor and right-handed solutions are excluded  at more than $2\sigma$ with the current data, while the HL-LHC will probe the two remaining scenarios in Tab.~\ref{tab:RDsFit}.                   

A caveat in this analysis concerns the range of convergence of the expansion in powers of $(s/\Lambda^2)$ implied by the EFT. This manifests, for instance, in the pathological behaviour of the cross section, Eq.~(\ref{eq:xsec_scheme}), for $\sqrt{s}\gg \Lambda$, leading to the upper bound $\Lambda\lesssim9$ TeV by means of unitarity arguments~\cite{DiLuzio:2017chi}. In the upper horizontal axis of Fig.~\ref{fig:EFTcb} we show the bounds in terms of the NP scale defined as $\Lambda=v/\sqrt{|V_{cb}||\epsilon_\Gamma|}$, which result to be within the range of $m_T$ reported by the experiments. The bins most sensitive to NP turn out to be those in 0.7 TeV $\lesssim m_T \lesssim$ 1.5 TeV; removing the tail of the distribution above that region has a minimal impact, of $\lesssim10\%$, on the bounds. Therefore, the EFT analysis should retain its validity for mediators above this scale.

For scenarios with lighter NP, the EFT study is invalid and one needs to do the analysis in terms of the particular UV completions of the operators. The possibilities in terms of mediators are also quite limited, reducing to the tree-level exchange of either new colorless vector ($W^\prime$)~\cite{Greljo:2015mma,Boucenna:2016wpr,Megias:2017ove,He:2017bft,Matsuzaki:2017bpp,Greljo:2018ogz,Asadi:2018wea} and scalar ($H^\pm$)~\cite{Tanaka:1994ay,Celis:2012dk,Celis:2016azn,Iguro:2017ysu,Fraser:2018aqj} particles in the $s$-channel, or leptoquarks in the $t$-channel~\cite{Sakaki:2013bfa,Alonso:2015sja,Barbieri:2015yvd,Freytsis:2015qca,Fajfer:2015ycq,Li:2016vvp,Barbieri:2016las,Becirevic:2016yqi,Crivellin:2017zlb,Cai:2017wry,Assad:2017iib,DiLuzio:2017vat,Bordone:2017bld,Barbieri:2017tuq,Marzocca:2018wcf,Greljo:2018tuh,Blanke:2018sro,Bordone:2018nbg,Becirevic:2018afm,DiLuzio:2018zxy,Angelescu:2018tyl,Heeck:2018ntp,Robinson:2018gza,Azatov:2018kzb}. We will not consider extra Higsses because they are in conflict with bounds from the decay $B_c\to\tau\nu$~\cite{Alonso:2016oyd,Akeroyd:2017mhr}. 


\textit{\textbf{The Leptoquark completion:}} Leptoquarks (LQ) carrying different quantum numbers (or combinations thereof) can produce all the operators in Eq.~(\ref{eq:leff1})~\cite{Sakaki:2013bfa,Alonso:2015sja,Barbieri:2015yvd,Freytsis:2015qca,Fajfer:2015ycq,Li:2016vvp,Barbieri:2016las,Becirevic:2016yqi,Crivellin:2017zlb,Cai:2017wry,Greljo:2018tuh,Assad:2017iib,DiLuzio:2017vat,Bordone:2017bld,Barbieri:2017tuq,Marzocca:2018wcf,Blanke:2018sro,Bordone:2018nbg,Becirevic:2018afm,DiLuzio:2018zxy,Angelescu:2018tyl,Heeck:2018ntp,Robinson:2018gza,Azatov:2018kzb} (we will use same notation as in refs.~\cite{Buchmuller:1986zs,Dorsner:2016wpm}). Our analysis involve $\textit{(i)}$ the scalar LQ $S_1=(\bar 3,1,1/3)$ producing vector-current (left-handed or right-handed) solutions; $\textit{(ii)}$ the $S_1$ producing the scalar-tensor solution; $\textit{(iii)}$ the $S_1$ combined with the scalar LQ $R_2=(3,2,7/6)$ to achieve a tensor solution by adjusting the masses $M_{S_1}= M_{R_2}$; $\textit{(iv)}$ the vector LQ $U_1=(3,1,2/3)$ leading also to the vector-current scenarios. All in all, we study four different LQ models, accounting for a total of six different NP solutions to the $R_{D^{(*)}}$ anomalies.   

We simulate the signals scanning the LQ masses in the range $0.75$ TeV to $5$ TeV and, for a given mass, we derive upper bounds on the product of LQ couplings to $c$- and $b$-quarks. In contrast to the EFT analysis, we simulate without jets at parton level in the final state keeping only the $t$-channel contributions, which are those connected to $R_{D^{(*)}}$. Single- and pair-LQ production topologies appear with extra jets. These introduce model dependence in terms of e.g. branching fractions to other possible decay channels, and are the target of direct searches (see e.g.~\cite{Diaz:2017lit,Dorsner:2018ynv}). 

In all the models we find that the bounds on the coupling-mass plane of the LQ are approximately equal to those derived from the EFT solutions they incarnate for masses $\gtrsim 2-3$ TeV. Solutions with lower masses are, nevertheless, being cornered by the aforementioned direct searches. Therefore, the conclusions for the LQ are very similar to the EFT analysis: The two LQ $S_1$-$R_2$ scenario is excluded by more than 2$\sigma$ in all the mass range. Right-handed solutions~\cite{Robinson:2018gza,Azatov:2018kzb} with $S_1$ and $U_1$ are also excluded by $\gtrsim2\sigma$ except for masses below 2 TeV. This mass range will be accessible with $\sim 150$ fb$^{-1}$ expected to be gathered after run 2 of the LHC. Finally, the left-handed ($S_1$ or $U_1$) and scalar-tensor ($S_1$) scenarios are not being probed yet but will be covered at the HL-LHC for almost the full mass range. We show in Fig.~\ref{fig:simplified_models}, left, a coupling-mass plot for the $U_1$ vector LQ illustrating our results and conclusions ($\mathcal{L} \supset g_{c}\, \bar c \gamma_\mu P_{L,R}\nu \, U_1^\mu + g_{b}\, \bar b \gamma_\mu P_{L,R}\tau \, U_1^\mu$). Similar plots for the other LQ have been presented elsewhere~\cite{AdmirTalk}.

\textit{\textbf{The $W'$ completion:}} The left-handed solution can be completed by a new massive spin-1 real $SU(2)_L$ triplet vector, $W'_L =({\bf 1},{\bf 3},0)$~\cite{Greljo:2015mma} (see also~\cite{Boucenna:2016qad,Barbieri:2016las,Megias:2017ove}). The neutral component of the triplet (a $Z'$ boson nearly degenerate to $W'^{\pm}$) leads to dangerous tree-level effects in neutral meson mixing. The flavour structure that keeps the contribution in $\Delta F =2$ observables under control, unavoidably predicts a $\mathcal{O}(V^{-1}_{c b})$ enhancement in $b \bar b \to Z' \to \tau^+ \tau^-$. Recast of the ATLAS $\tau^+ \tau^-$ search with $3.2$~fb$^{-1}$ at 13~TeV, performed in Ref.~\cite{Faroughy:2016osc}, already cuts deep into the model's perturbative parameter space explaining the anomaly, requiring the $Z'$ to be a rather wide resonance. A second class of models involve a complex vector, $SU(2)_L$ singlet with a hypercharge, $W^\prime_R=({\bf 1},{\bf 1},+1)$, and a relatively light right-handed neutrino that induces the right-handed solution to $R_{D^{(*)}}$~\cite{Greljo:2018ogz,Asadi:2018wea} (see also~\cite{Carena:2018cow,Babu:2018vrl}). Explicit UV models introduce a $Z'$ boson with flavor violating effects completely decoupled from $R_{D^{(*)}}$ due to the lack of $SU(2)_L$ relations.

The relevant $W'$ interactions are defined as, $\mathcal{L} \supset g_{b c}\, \bar c \gamma^\mu P_{L,R}b \, W'_\mu + g_{\tau \nu}\, \bar \nu \gamma^\mu P_{L,R} \tau \, W'_\mu$\,+\,h.c., where the chirality is inaccessible in our present analysis. We perform simulations using the same specifications as for the EFT, for several $W'$ masses in the range 0.5~TeV to 3.5~TeV and different total width hypotheses. Besides including experimental systematics and the SM theory uncertainties, we also estimate the uncertainty on the signal prediction stemming from the higher-order QCD corrections and PDF determination. These uncertainties combined in quadrature range from roughly $10\%$ ($30\%$) for $m_{W'} = 1$~TeV (3~TeV). For a given mass and width combination, we set an upper limit on the product of the two couplings in the $W'bc$ and $W'\tau\nu$ vertices, and confront it with the fit results from $R_{D^{(*)}}$. Note that this procedure is rather general, and it does not require to specify any additional $W'$ decay modes. This choice of parameters is suitable for the interpretation of the perturbativity of the model. Very wide resonances indicate the loss of predictivity and here we investigate up to $\Gamma_{W'} \lesssim 0.5\, M_{W'}$. Our results, shown in Fig.~\ref{fig:simplified_models}~(right) {in solid (dashed) for observed (expected)}, exclude the $W'_R$ models in the pertubatively calculable parameter space explaining the anomaly, $|g_{b c} g^{*}_{\tau \nu}| / M^2_{W'} \approx (0.6\pm0.1)~\rm{TeV}^{-2}$. This quantity is $\approx (0.14\pm0.03)~\rm{TeV}^{-2}$ for the left handed solution, which is, however, scrutinised by the $Z' \to \tau^+ \tau^-$ searches at the LHC~\cite{Faroughy:2016osc}.

A potential caveat could be the loss of sensitivity in the low $W'$ mass region as the signal tends to hide in the large SM background. Robust lower limits of $\gtrsim 100$~GeV on a new electrically-charged gauge boson from the LEP experiment are significantly improved by the electroweak $p p \to W'^+ W'^-$ pair-production process at the LHC~\cite{JAJfuture}. Another promising direction to close this window is to study $p p \to \tau \nu$ searches at previous $p p$ collision energies~\cite{JAJfuture}. Search strategies in this region could include requiring a $b$-tag in the final jets~\cite{Abdullah:2018ets}. Some sensitivity is expected also in the top quark decays~\cite{Kamenik:2018nxv}.


\textit{\textbf{The semi-tauonic $b\to u$ transitions:}} NP models addressing $R_{D^{(*)}}$ are expected to contribute to semi-tauonic transitions other than $b\to c$, and to neutral-current processes via $SU(2)_L$ symmetry (e.g. for the LQ or $W^\prime_L$). Focusing on the charged-currents and their impact on the mono-tau signal at the LHC, we conclude from Eq.~(\ref{eq:xsec_scheme}), that additional flavor structures can only enhance the $p p \to \tau  \nu$ signal~\footnote{With the exception of contributions in the valence $ud$ flavor entry that we do not consider because it is tightly constrained by $\tau$ decays~\cite{Cirigliano:2018dyk}.}. Thus, the bounds obtained above are \textit{conservative} in the sense that they can only be stronger in realistic models of NP. 

\begin{table}[h]
\caption{$2\sigma$ upper bounds for the absolute value of the WCs of semi-tauonic $ub$ transitions at $\mu=m_b$.~\label{tab:fitsub}}
\begin{tabular}{cccc}
\hline\hline
Data set& Vector& Scalar & Tensor\\
\hline
LHC combined          & $0.72$ & $1.23$ & $0.34$\\
\hline
LHC (150 fb$^{-1}$)       &$0.48$ & $0.84$ & $0.23$\\
HL-LHC  & $0.21$ & $0.37$ & $0.10$\\
\hline\hline
\end{tabular}
\end{table}

We explore this issue by repeating our analysis for $b\to u$ operators in the EFT. These are transitions particularly interesting because they are typically affected by NP addressing $R_{D^{(*)}}$. Experimentally, branching fractions of $B\to \tau\nu$ have been measured, showing a slight excess over the SM at $\sim 1.5\sigma$, while there is only an upper limit on the semi-tauonic decay $B^0\to\pi^-\tau^+\nu$. In Tab.~\ref{tab:fitsub}, we show the bounds on the different structures that are obtained from $pp\to\tau\nu X$ at the LHC, assuming that these are the only active flavor entries. The limits on the $ub$ WCs are roughly a factor two worse than for the $cb$ ones, which is the result of the CKM suppression $(|V_{ub}|/|V_{cb}|)^2$ partially compensated by the larger PDFs of the $up$-quark, \textit{c.f.}  Eq.~(\ref{eq:xsec_scheme}). Nonetheless, these are competitive with those obtained from $B$ decays. {In particular, LHC bounds are currently better than the ones derived from $B^0\to\pi^-\tau^+\nu$~\cite{Hamer:2015jsa}, $-1.25\lesssim\epsilon_T^{ub}\lesssim0.57$ and  $-1.75\lesssim\epsilon_{S_L}^{ub}+\epsilon_{S_R}^{ub}\lesssim 0.94$ at 2$\sigma$, using the form factors from lattice QCD calculations~\cite{Lattice:2015tia,Bailey:2015nbd}.}  

%

\textit{\textbf{Conclusions and discussion:}} We have discussed in detail the consequences of the univocal connection between the semi-tauonic $B$ decays and the $pp\to\tau_h X+{\rm MET}$ signature at the LHC given by crossing symmetry, \textit{cf.} Fig.~\ref{fig:intro}. Our key findings can be summarized as follows: \textit{\textbf{First}}, the current data at 13 TeV on $W^\prime$ searches, consisting of roughly $\sim36$ fb$^{-1}$ per collaboration, is already sensitive to NP scenarios addressing the $R_{D^{(*)}}$ anomalies. Pure tensor solutions, completed by LQ, and right-handed solutions, completed by $W^\prime_R$ or LQ, are excluded at more than 2$\sigma$ for most of masses. \textit{\textbf{Second,}} the sensitivity that will be achieved by extrapolating through the HL-LHC phase will probe \textit{all} the possible scenarios that explain the anomalies. Therefore mono-tau searches can provide a ``no-loose theorem'' or ``the ultimate test'' for the confirmation of such NP at the LHC. \textit{\textbf{Third,}} the LHC is also competitive in the $b\to u$ transitions and bounds on some NP scenarios are currently better than in $B$ decays. This illustrates the impact, and complementarity with low-energy experiments, that a program of high-precision measurements at the LHC can have in Flavor Physics. 

{In our analysis, the sensitivity to NP comes mainly from the $m_T$ bins around $\sim 1$ TeV, while the EFT provides a {good} description of explicit models (with the exception of light $W^\prime$). The implied constraints are difficult to avoid by more elaborate model building (compared to e.g.~\cite{Faroughy:2016osc}).  Finally, significant improvements of the present analysis are possible in the future. For instance, exploiting $\tau_h$ charge-asymetries, rapidity distribution and polarization could help improving the signal over background discrimination. 
 Another avenue would be to consider adding data from the leptonic tau decays. A detailed study of these aspects and their impact on the sensitivity will be presented elsewhere~\cite{JAJfuture}. }

\section*{Acknowledgements}
 
We thank Martin Gonzalez-Alonso, David Marzocca, and Christos Vergis for useful discussions. JMC acknowledges support from the Spanish MINECO through the ``Ram\'on y Cajal'' program RYC-2016-20672.

\bibliographystyle{apsrev4-1}

\bibliography{RDincoll.bib}

\end{document}

%% file: definitions_notes.tex
\newcommand{\fref}[1]{Fig.~\ref{fig:#1}} 
\newcommand{\eref}[1]{Eq.~\eqref{eq:#1}} 
\newcommand{\erefn}[1]{ (\ref{eq:#1})}
\newcommand{\erefs}[2]{Eqs.~(\ref{eq:#1}) - (\ref{eq:#2}) } 
\newcommand{\aref}[1]{Appendix~\ref{app:#1}}
\newcommand{\sref}[1]{Section~\ref{sec:#1}}
\newcommand{\cref}[1]{Chapter~\ref{ch:.#1}}
\newcommand{\tref}[1]{Table~\ref{tab:#1}}

\newcommand{\nn}{\nonumber \\}  
\newcommand{\nnl}{\nonumber \\}  
\newcommand{\nl}{& \nonumber \\ &}
\newcommand{\bnl}{\right .  \nonumber \\  \left .}
\newcommand{\dbnl}{\right .\right . & \nonumber \\ & \left .\left .}

\newcommand{\beq}{\begin{equation}} 
\newcommand{\eeq}{\end{equation}} 
\newcommand{\ba}{\begin{array}}  
\newcommand{\ea}{\end{array}} 
\newcommand{\bea}{\begin{eqnarray}}  
\newcommand{\eea}{\end{eqnarray} }  
\newcommand{\be}{\begin{eqnarray}}  
\newcommand{\ee}{\end{eqnarray} }  
\newcommand{\bal}{\begin{align}}
\newcommand{\eal}{\end{align}}   
\newcommand{\bi}{\begin{itemize}}  
\newcommand{\ei}{\end{itemize}}  
\newcommand{\ben}{\begin{enumerate}}  
\newcommand{\een}{\end{enumerate}}  
\newcommand{\bc}{\begin{center}}
\newcommand{\ec}{\end{center}} 
\newcommand{\bt}{\begin{table}}
\newcommand{\et}{\end{table}}  
\newcommand{\btb}{\begin{tabular}}
\newcommand{\etb}{\end{tabular}}  
\newcommand{\bvec}{\left ( \ba{c}}
\newcommand{\evec}{\ea \right )}

\newcommand{\cO}{{\mathcal O}} 
\newcommand{\co}{{\mathcal O}} 
\newcommand{\cL}{{\mathcal L}} 
\newcommand{\cl}{{\mathcal L}} 
\newcommand{\cM}{{\mathcal M}}

\newcommand{\const}{\mathrm{const}}

\newcommand{\ev}{ \mathrm{eV}}
\newcommand{\kev}{\mathrm{keV}}
\newcommand{\mev}{\mathrm{MeV}}
\newcommand{\gev}{\mathrm{GeV}}
\newcommand{\tev}{\mathrm{TeV}}

\newcommand{\mpl}{M_{\mathrm Pl}}

\def\mgut{\, M_{\rm GUT}}
\def\tgut{\, t_{\rm GUT}}
\def\mpl{\, M_{\rm Pl}}
\def\mkk{\, M_{\rm KK}}
\newcommand{\msusy}{M_{\rm soft}}

\newcommand{\dslash}[1]{#1 \! \! \! {\bf /}}
\newcommand{\ddslash}[1]{#1 \! \! \! \!  {\bf /}}

\def\ads{AdS$_5$\,}
\def\adse{AdS$_5$}
\def\intdk{\int {d^4 k \over (2 \pi)^4}} 

\def\ra{\rangle}
\def\la{\langle}  

\def\sgn{{\rm sgn}}
\def\pa{\partial}  
\newcommand{\dlr}{\overleftrightarrow{\partial}}
\newcommand{\Dlr}{\overleftrightarrow{D}}
\newcommand{\re}{{\mathrm{Re}} \,}
\newcommand{\im}{{\mathrm{Im}} \,}
\newcommand{\tr}{\mathrm T \mathrm r}  

\newcommand{\Ra}{\Rightarrow}
\newcommand{\lra}{\leftrightarrow}
\newcommand{\llra}{\longleftrightarrow}

\newcommand\simlt{\stackrel{<}{{}_\sim}}
\newcommand\simgt{\stackrel{>}{{}_\sim}}   
\newcommand{\zt}{$\mathbb Z_2$ }

\newcommand{\ha}{{\hat a}}
\newcommand{\hab}{{\hat b}}
\newcommand{\hac}{{\hat c}} 

\newcommand{\ti}{\tilde}  
\def\hc{{\rm h.c.}} 
\def\ov{\overline}  
  
\newcommand{\eps}{\epsilon}

\def\cog{\color{OliveGreen}}
\def\cor{\color{Red}}
\def\copu{\color{purple}}
\def\coro{\color{RedOrange}}
\def\coma{\color{Maroon}}
\def\cob{\color{Blue}}
\def\cobr{\color{Brown}}
\def\cobl{\color{Black}}
\def\cost{\color{WildStrawberry}}

\newcommand{\eL}{\epsilon_L}
\newcommand{\eR}{\epsilon_R}
\newcommand{\eSL}{\epsilon_{S_L}}
\newcommand{\eSR}{\epsilon_{S_R}}
\newcommand{\eT}{\epsilon_T}
\def\slashp{p \!\!\! \slash}
\def\slashs{s \!\!\! \slash}
\def\slashE{E \!\!\! \slash_T}